\documentclass{emulateapj}
\usepackage{graphicx}
\usepackage{multirow,natbib}
\usepackage{textcomp}
\usepackage{epsfig}
\usepackage{amsmath}
\usepackage{amssymb}
\usepackage{amsthm}
\usepackage{xy}
\usepackage{pbox}

\begin{document}

\title{Variation of Spectral and Timing Properties in the Extended Burst Tails from the Magnetar 4U 0142+61}
\author{Manoneeta Chakraborty, Ersin G\"o\u{g}\"u\c{s}, Sinem \c{S}a\c{s}maz Mu\c{s}, Yuki Kaneko}

\affil{Sabanc\i~ University, Faculty of Engineering and Natural 
  Sciences, Orhanl\i ~Tuzla 34956 Istanbul Turkey}
  
\email{manoneeta@sabanciuniv.edu}

\begin{abstract}\label{Abstract}
Extended emission episodes with intensity above the pre-burst level are observed following magnetar bursts from a number of soft gamma repeaters (SGRs) and anomalous X-ray pulsars (AXPs). Such extended tail emission were observed subsequent to two events detected from AXP 4U 0142+61. We investigated in detail the evolution of spectral and temporal properties during these two tail segments using RXTE/PCA observations, and report distinct variations both in the spectral and temporal behavior throughout the tails. In particular, sudden enhancement of pulsation amplitude in conjunction with bursts, and smooth decline of X-ray emission (cooling) during the tail were observed in both cases. We suggest that an inefficiently radiating trapped fireball formed during the burst, which can heat up the stellar surface, is able to explain the tail properties and its energetics. We also present the episodic detection of absorption and emission features during tails. One possible mechanism that has been proposed to give rise to such spectral lines is the proton/ion cyclotron resonance process which has been suggested to offer a valuable tool in probing the complex magnetic field of magnetars.

\end{abstract}


\section{Introduction}\label{Introduction}

A small group of isolated neutron stars possessing exceptionally strong surface magnetic fields in the range $10^{14}-10^{15}$ G \citep{Duncan1992} are termed magnetars. These systems have relatively long spin periods ranging between 0.3 - 12 s, and emit X-rays in the luminosity range of $10^{33}-10^{36}$ ergs s$^{-1}$. 
Soft gamma repeaters (SGRs) are a sub-class of magnetars which are characterized by emitting high intensity and repeating bursts in the soft gamma ray band. Such bursts have peak luminosities of $10^{37}$ to $10^{41}$ erg s$^{-1}$, lasting typically for a duration of $\sim 0.1$ s \citep{Gogus2001}. In their active phases, SGRs can emit a few to sometimes thousands of bursts. From energetics and morphological perspectives, SGR bursts are broadly divided into three categories: The typical short bursts have isotropic energies upto $10^{41}$ erg, last for a small fraction of a second \citep{Gogus2001}. The intermediate events are more energetic, having isotropic energies of $10^{41}-10^{43}$ erg and lasting for few seconds \citep{Ibrahim2001, Mereghetti2009, Gogus2011}. The giant flares are extremely energetic events emitting isotropic energies of $10^{44}-10^{46}$ erg and lasting for hundreds of seconds \citep{Mazets1979, Hurley1999, Hurley2005}. Magnetar bursts at all scales are generally explained by the magnetar model \citep{Duncan1992} which suggests that the fracturing of the neutron star crust at rather localized or global scale, releasing build-up magnetic stress in the solid crust gives rise to these sudden flashes of energy release \citep{Thompson1995}. Alternatively, such bursts can also be explained by magnetic reconnection events in the highly magnetized neutron star environments \citep{Lyutikov2003, Lyutikov2015}.

Anomalous X-ray pulsars (AXPs) are also members of the magnetar family, and exhibit SGR like bursts (see for e.g., \citealt{Gavriil2002, Kaspi2003}). The AXP bursts occur usually less frequent, and energetically similar or a bit weaker than typical short SGR bursts \citep{Gavriil2004}. Of the 12 confirmed AXPs, six sources have showed SGR like bursts (see \citet{Olausen2014} and the online list of magnetars\footnote{http://www.physics.mcgill.ca/$\sim$pulsar/magnetar/main.html} for their bursting behavior). 1E 2259+586 is currently the most prolific AXP in bursting; 80 bursts were detected in a single \textit{Rossi X-ray Timing Explorer (RXTE)} observation in 2002 \citep{Gavriil2004}. Although these bursts resemble SGR bursts in many respects, there were some notable differences, for example, the more energetic AXP bursts have the harder spectra while SGR burst spectra tend to get softer with increasing energy \citep{Gavriil2004}. 

The group of intermediate energy events emerged with leading bursts followed by prolonged emission episodes with intensities clearly above the pre-burst emission level. These so called bursts with extended tails have been detected from a number of magnetars;- SGR 1900+14 \citep{Ibrahim2001, Lenters2003}, SGR 1806--20 \citep{Gogus2011}, SGR J1550--5418 \citep{Kuiper2012, SasmazMus2015}. The extended tails last for about few tens of seconds to thousands of seconds. The energy emitted during such extended tails is a small fraction of the total energy of the leading bursts \citep{Lenters2003,Gogus2011}. However, recent investigations of SGR 1550-5418 bursts with extended tails indicated that energetics of the tails are not always linked; many bright bursts do not exhibit extended tails whereas those observed tails were preceded by much lower intensity bursts \citep{SasmazMus2015}. Detailed studies of the extended tails from all these sources yielded distinct timing and spectral evolution properties during these phases \citep{Ibrahim2001, Lenters2003, SasmazMus2015}. The tail spectra are generally described with a thermal shape (using a black body function with temperature, 1.5-4 keV) and exhibit cooling trend. This can be explained by the cooling of thermal component on the neutron star surface which was heated up by the preceding burst. 

4U 0142+61 is currently the brightest persistent AXP source (see table 2 of online magnetar catalog$^1$ and references therein; \citealt{Patel2003}). This source is spinning with a period of 8.7 s and has period derivative of $0.2\times10^{-11}$ s s$^{-1}$ which implies a surface dipole magnetic field of strength $1.3\times10^{14}$ G. \textit{RXTE} monitored this source for more than a decade. The spin frequency of this source generally showed high degree of stability \citep{Gavriil2002, Dib2007} though there have been some indications of a glitch \citep{Morii2005}. This source entered an active phase in 2006 March, when remarkable changes in temporal and spectral properties were observed, including several bursts from this particular AXP were observed for the first time \citep{Gavriil2011}. \citet{Gavriil2011} studied the spectral and temporal characteristics of these bursts. The T$_{90}$ durations \citep{Kouveliotou1993} of these bursts spanned a very wide range from 0.4 s to 1757 s. For each burst, they extracted a single X-ray spectrum considering its entire T$_{90}$ interval. They found that the spectra of five bursts were well represented with single blackbody model, while they reported unusual signatures of spectral emission features on top of the single component blackbody model for the sixth burst. They concluded that a broad spectral feature at $\sim14$ keV along with a narrow feature at $\sim 8$ keV were additionally required to explain the spectrum well. 
They suggested time-dependent spectral variations by employing an unusual time-resolved spectral investigation technique (see Figure 4 in \citealt{Gavriil2011}). They showed that the pulse profiles of the source in the observations comprising the bursts exhibit significant changes, showing additional pulse peaks. The 4-20 keV pulsed flux during the decay of the bursts was increased following the bursts \citep{Gavriil2011}. They also performed long term timing investigation and provided a phase coherent timing solution, and uncovered a likely anti-glitch (sudden spin-down event) in conjunction with active phase.	
  	
Motivated by the indications of the spectral and temporal variations during the tails of these bursts and to better understand the nature of reported spectral features, we investigated the bursts from 4U 0142+61 particularly focusing on the extended tail emission episodes. We describe the \textit{RXTE} data used in section 2 and explain the method of analyzing the extended tails in section 3. In section 3, we present the results of our time resolved spectral and timing analyses during the tails. The physical implications of our results and comparison with previous observations are then discussed in the final section.

\section{Observation}\label{Observation}
\begin{figure}[H!ta]
\includegraphics[width=8cm]{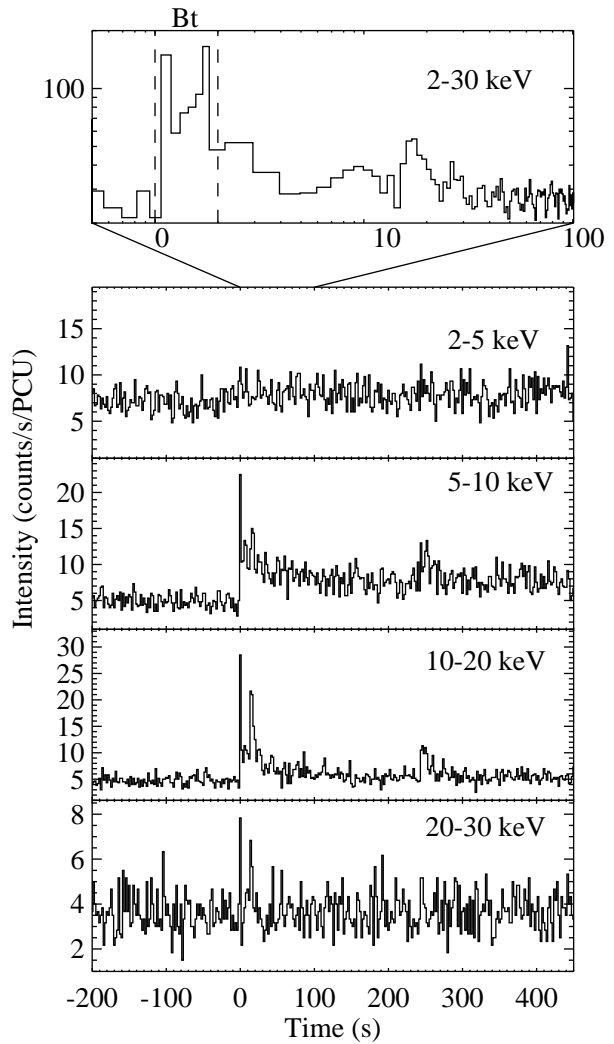}
\caption{Light curve of event A over different energy bands. The topmost zoomed-up panel shows the burst lightcurve starting from 0.5 s before the burst to 100 s after in 2.0-30.0 keV energy band. The segment marked `Bt' denotes the time interval corresponding to the burst for event A. \label{Alc}}
\end{figure}

The \textit{RXTE} routinely monitored 4U 0142+61 starting from 2000 until the end of the mission at the end of 2011. Here, we only focused on the observations comprising the bursts reported by \citet{Gavriil2011}, with data collected with the Proportional Counter Array (PCA) \citep{Jahoda2006} on-board \textit{RXTE}. The PCA consisted of 5 proportional counter units (PCUs) each of which was made up of one Propane veto, 3 Xenon and one Xenon veto layers. The PCA operated in the energy range 2-60 keV and has a large effective area of $\sim 6500$ cm$^{2}$ which made it a highly sensitive instrument. For our analysis, we used data in Good-Xenon mode as it offers an excellent time resolution of 1 $\mu$s and 256 channel spectral capability. The data was analyzed using the relevant HEASOFT v6.16 tools and the CALDB version 20120110. For timing analysis, we converted the photon arrival times to that at the solar system barycenter. IDL version 8.5 was used extensively for the purposes of timing analyses and plotting. For this analysis, we concentrated on the bursts on 2006 June 25 (bursts 2, 3, 4, 5 in \citealt{Gavriil2011}) which we refer as event A, and the burst on 2007 February 07 (burst 6 in \citealt{Gavriil2011}) referred as event B in the remainder of this paper.

\begin{figure}[H!tba]
\includegraphics[width=8cm]{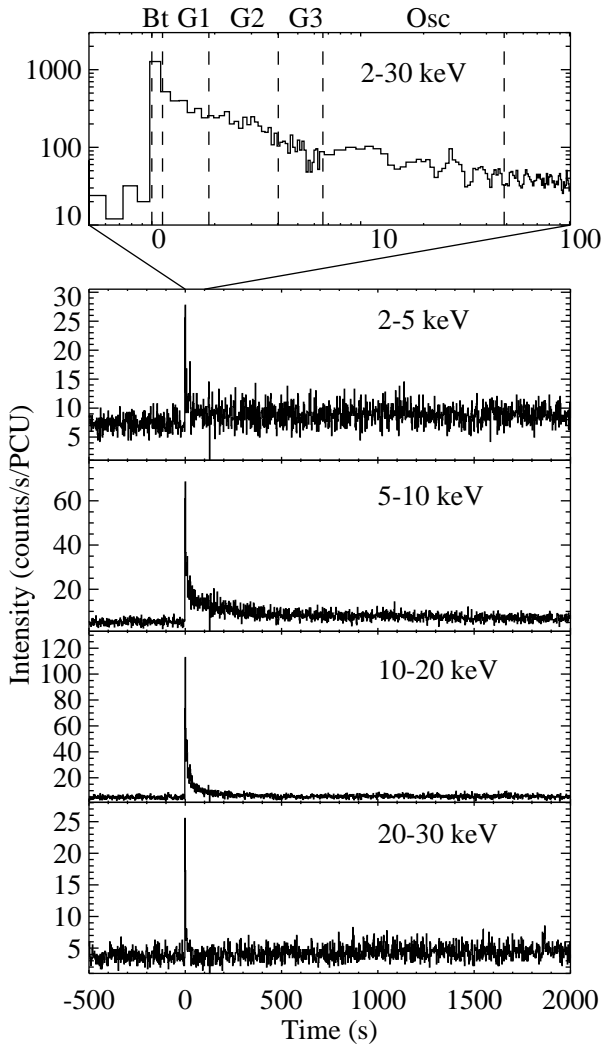}
\caption{Light curve of event B over different energy bands. The topmost zoomed-up panel shows the burst lightcurve starting from 0.5 s before the burst to 100 s after in 2.0-30.0 keV energy band. The segment marked `Bt' denotes the time interval corresponding to the burst for event B. The segments marked out `G1', `G2', `G3' and `Osc' denotes the Gray 1, Gray 2, Gray 3 and Oscillation regions respectively. Periodic pulsation, clearly observable by eye, can be seen in the oscillation and following region in the top panel. \label{Blc}}
\end{figure}

\section{Analysis of extended tails}\label{Tail}

We identified the extended tail as the interval following a burst during which the emission is clearly greater than the pre-burst level. The extended tails following the bursts in event A and event B continue until the end of the observing window and correspond to duration of 463 and 1600 s respectively. Figures \ref{Alc} and \ref{Blc} show the light curves for the events A and B respectively. 
The bottom four panels of each of the two plots show the energy resolved light curves for the energy ranges 2.0-5.0, 5.0-10.0, 10.0-20.0 and 20.0-30.0 keV. These light curves are binned with 2 s time resolution for illustration purposes. However, burst identifications and tail segmentations were done with much finer time resolution as we discuss below. The top panels of each plot show the zoomed-in light curve of the bursts spanning an interval between 0.5 s before the burst and 100 s after. Periodic modulations at the spin period can be observed during the tail of both events, especially in the zoomed-in panels. 
We investigated the spectral evolution of events with extended emission by separating them into different time segments: We inspected the light curves of these events using different time resolutions and have seen clear distinct features in 31.25 ms binned light curves. We used these features and pulse amplitude variations as a guideline to separate them into different segments. Finally, we fine tuned the borders between the regions, especially during the first $\sim$100 s of events, by inspecting the variations in the spectra and divided event A in ten different regions and event B in thirteen different regions. The events start with a significantly high intensity peak which we name as burst and gradually decaying enhanced emission region. We do not exclude any possible weak burst features observed at 31.25 ms time resolution during the extended tail as they could not be significantly distinguished from noise and also did not perceptibly affect the properties of the tail segments. Although the light curve of event A is fairly simple (a distinct burst + segments with enhanced emission), light curve of event B consists of morphologically and spectrally different regions which we additionally named as gray and oscillation regions. Gray regions are the transition interval between the burst and oscillation phases. Oscillation region follows the gray zone and has clearly seen oscillations at the spin frequency of the source (see the top panel of Figure~\ref{Blc}). The durations and the corresponding counts of each segment during these two events are exhibited in Table~\ref{cnttable}. The total count given in each segment is background subtracted.

\begin{table}[H!tba]
\centering
\caption{Table displaying the details of each segment corresponding to the two events from 4U 0142+61 \label{cnttable}}
\begin{tabular}{cccccccc}
\hline
Event & Observation & Segment & Total$^a$ & Segment$^b$  & Active\\
      & date        &         & count     & duration (s) &  PCUs\\
\hline
{\bf A} & {\bf 2006 June 25}    &         &       &       &  0, 2, 3\\
  &                 & Burst   &  191.29    &    1.0       & \\
  &                 & Tail 1  & 1780.57    &    50.0      & \\ 
  &                 & Tail 2  &  947.57    &    50.0      & \\
  &                 & Tail 3  &  737.57    &    50.0      & \\
  &                 & Tail 4  &  509.57    &    50.0      & \\
  &                 & Tail 5  &  469.82    &    50.0      & \\
  &                 & Tail 6  & 1035.57    &    50.0      & \\
  &                 & Tail 7  &  530.57    &    50.0      & \\
  &                 & Tail 8  &  754.57    &    50.0      & \\
  &                 & Tail 9  &  791.82    &    50.0      & \\
\hline
{\bf B} & {\bf 2007 Feb 07}     &         &        &      & 0, 2\\
  &                 & Burst   &   254.15    &    0.125    & \\
  &                 & Gray 1  &   535.91    &     0.75    & \\
  &                 & Gray 2  &   789.37    &     2.16    & \\  
  &                 & Gray 3  &   360.62    &     2.56    & \\
  &                 & Osc     &  3201.84    &    42.06    & \\    
  &                 & Tail 1  &  4993.12    &    200.0    & \\ 
  &                 & Tail 2  &  3069.12    &    200.0    & \\
  &                 & Tail 3  &  2161.12    &    200.0    & \\
  &                 & Tail 4  &  2147.12    &    200.0    & \\
  &                 & Tail 5  &  2544.12    &    200.0    & \\
  &                 & Tail 6  &  2396.12    &    200.0    & \\
  &                 & Tail 7  &  1985.12    &    200.0    & \\
  &                 & Tail 8  &  1748.12    &    200.0    & \\  
\hline  
\end{tabular}
\begin{footnotesize}
\begin{flushleft}
$^a$ Background subtracted total count obtained for each spectrum \\
$^b$ Duration as considered for spectral and timing analysis (see text)
\end{flushleft}
\end{footnotesize}
\end{table}

\subsection{Spectral Analysis}\label{Spec}
We extensively investigated the variation of spectral properties throughout the extended tail emission episodes, particularly on short timescales, to probe any changes in its continuum and the presence of line features. The spectrum for each segment during the tail was extracted and spectral fits were performed in the energy range 2.5-30 keV using XSPEC software version 12.9. We used an interstellar neutral hydrogen column density of $7.3\times10^{21}$ cm$^{-2}$ \citep{Weng2015} which was kept constant in spectral fits. The background spectra were extracted from 350 s and 438 s long data segments in the pre-burst data of event A and event B respectively. Each spectrum was accumulated from all layers of operating PCUs, and then grouped in such a way so that each spectral bin would include at least 20 counts. We fitted continuum spectra using a variety of thermal and non-thermal models, as well as their combinations. We present below the details of our investigations for both events A and B.

\subsubsection{Event A}
\begin{figure}[H!ba]
\centering
\includegraphics[width=8cm]{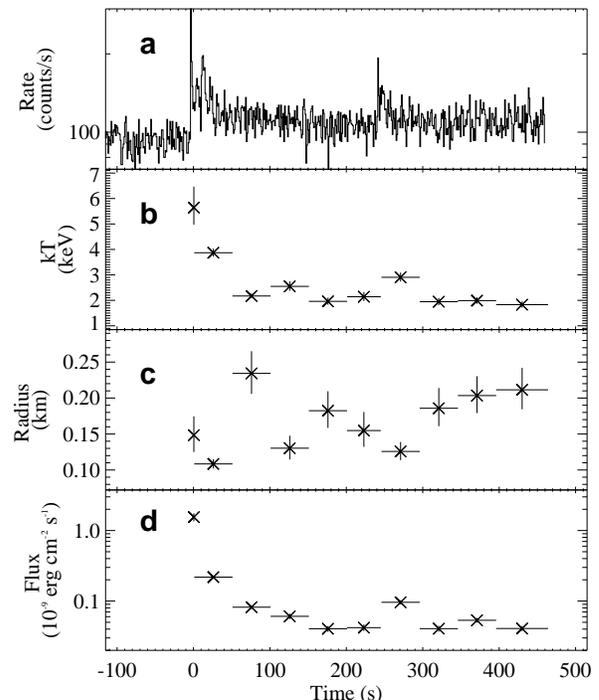}
\vspace*{-1.0cm}
\caption{Variation of best-fit spectral parameters of the burst and extended tail in event A. (a) The source intensity of the event with 1 s binning. (b) The evolution of the blackbody temperature. (c) The radius of the corresponding black body emitting region. (d) The variation of the 2.5-30 keV unabsorbed flux during the burst and the tail segments. All errors shown here are $1 \sigma$ errors on the parameters (the errors on the flux values are relatively too small to discern in the plot).\label{Aspecvar}}
\end{figure}

\begin{table*}[H!tba]
\centering
\caption{Table displaying the best-fit spectral parameters for event A \label{SpecresultsA}}
\begin{tabular}{ccccccccc}
\hline
Segment & kT & Radius & \multicolumn{2}{c}{Absorption 1}& \multicolumn{2}{c}{Absorption 2} & Flux$^a$ & $\chi_{\nu}^2$ (dof)\\
     & (keV) & (km) & E (keV) & Norm & E (keV) & Norm &$10^{-9}$ ergs/s/cm$^2$ & \\
\hline
Burst &  5.64$_{- 0.67}^{+ 0.83}$ &  0.15$_{- 0.02}^{+ 0.03}$ & ... & ... & ... & ... &  1.550$_{- 0.199}^{+ 0.205}$ &  0.53 (10)\\
Tail 1 &  3.87$_{- 0.15}^{+ 0.16}$ &  0.11$_{- 0.01}^{+ 0.01}$ & ... & ... & ... & ... &  0.218$_{- 0.011}^{+ 0.012}$ &  1.31 (58)\\
Tail 2 &  2.17$_{- 0.12}^{+ 0.13}$ &  0.23$_{- 0.03}^{+ 0.03}$ & 4.56$_{-0.13}^{+0.13}$ & -0.00139$_{- 0.00037}^{+ 0.00036}$ & 6.70$_{-0.17}^{+0.15}$ & -0.00089$_{- 0.00022}^{+ 0.00022}$ &  0.082$_{- 0.005}^{+ 0.005}$ & 0.66 (56)\\
Tail 3 &  2.55$_{- 0.18}^{+ 0.20}$ &  0.13$_{- 0.02}^{+ 0.02}$ & ... & ... & ... & ... &  0.061$_{- 0.005}^{+ 0.006}$ &  1.21 (61)\\
Tail 4 &  1.96$_{- 0.14}^{+ 0.15}$ &  0.18$_{- 0.02}^{+ 0.03}$ & ... & ... & ... & ... &  0.040$_{- 0.004}^{+ 0.004}$ &  1.53 (57)\\
Tail 5 &  2.14$_{- 0.17}^{+ 0.19}$ &  0.15$_{- 0.02}^{+ 0.03}$ & ... & ... & ... & ... &  0.042$_{- 0.004}^{+ 0.005}$ &  1.29 (53)\\
Tail 6 &  2.91$_{- 0.17}^{+ 0.18}$ &  0.13$_{- 0.01}^{+ 0.01}$ & ... & ... & ... & ... &  0.096$_{- 0.007}^{+ 0.007}$ &  0.85 (60)\\
Tail 7 &  1.95$_{- 0.14}^{+ 0.15}$ &  0.19$_{- 0.02}^{+ 0.03}$ & ... & ... & ... & ... &  0.041$_{- 0.004}^{+ 0.004}$ &  1.29 (58)\\
Tail 8 &  1.99$_{- 0.12}^{+ 0.14}$ &  0.20$_{- 0.02}^{+ 0.03}$ & ... & ... & ... & ... &  0.053$_{- 0.004}^{+ 0.004}$ &  0.69 (62)\\
Tail 9 &  1.83$_{- 0.12}^{+ 0.13}$ &  0.21$_{- 0.03}^{+ 0.03}$ & ... & ... & ... & ... &  0.041$_{- 0.003}^{+ 0.003}$ &  1.12 (63)\\
\hline
\end{tabular}
\begin{footnotesize}
\\
\begin{flushleft}
$^a$ Unabsorbed flux in 2.5-30 keV in units of $10^{-9}$ ergs/s/cm$^2$.\\
All parameter errors quoted here $1 \sigma$ confidence errors.	
\end{flushleft}
\end{footnotesize}
\end{table*}

\begin{figure*}[H!tba]
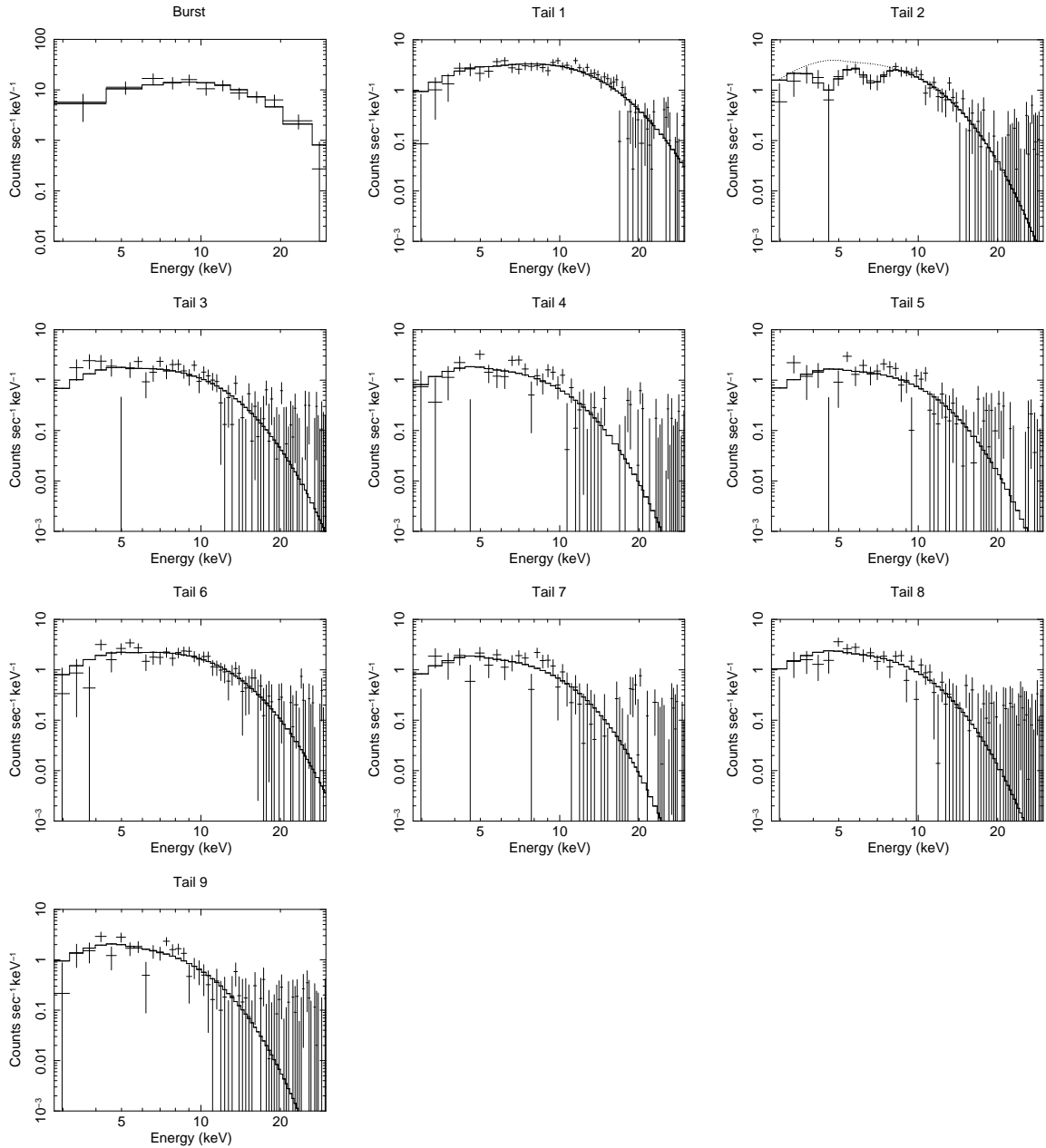

\centering
\begin{tabular}{ccc}
\includegraphics[width=4cm, angle=-90]{fig4_1.ps} &
\includegraphics[width=4cm, angle=-90]{fig4_2.ps} &
\includegraphics[width=4cm, angle=-90]{fig4_3.ps} \\
\includegraphics[width=4cm, angle=-90]{fig4_4.ps} &
\includegraphics[width=4cm, angle=-90]{fig4_5.ps} &
\includegraphics[width=4cm, angle=-90]{fig4_6.ps} \\
\includegraphics[width=4cm, angle=-90]{fig4_7.ps} &
\includegraphics[width=4cm, angle=-90]{fig4_8.ps} &
\includegraphics[width=4cm, angle=-90]{fig4_9.ps} \\
\includegraphics[width=4cm, angle=-90]{fig4_10.ps} 
\end{tabular}
\caption{Spectral fit of segments corresponding to event A. The label at the top of each panel denotes the segment that the corresponding spectrum was generated from. Each panel also show the best fit spectral model in the 2.5-30 keV band. \label{Aspecfits}}
\end{figure*}

We divided the $\sim460$ s long extended tail emission into the segments of 50 s in order to perform time-resolved spectral analysis. We provide the details of the spectral analysis results in Table~\ref{SpecresultsA} and present time evolution of model parameters in Figure~\ref{Aspecvar}. The burst spectrum was fitted well with a single black body model (Figure~\ref{Aspecfits}), temperature kT = 5.64$_{- 0.67}^{+ 0.83}$ keV ($\chi^2_{\nu} = 0.53$ for 10 degrees of freedom, dof). A single non-thermal model (power law) yielded a worse fit giving a $\chi^2_{\nu} =1.10$ for the same dof. Using the black body fit results, we obtained the unabsorbed burst flux in the 2.5-30 keV as $1.6 \pm 0.2 \times 10^{-9}$ erg s$^{-1}$ cm$^{-2}$, which corresponds to a total burst luminosity of $2.3 \times 10^{36}$ erg s$^{-1}$ (assuming the distance to the source as 3.5 kpc \citep{Durant2006}). The tail spectra (except for the second tail segment) were also best described by a single black body with the temperature showing a smooth decline trend: the black body temperature was 3.87$_{- 0.15}^{+ 0.16}$ keV just following the burst and it decreased to 1.83$_{- 0.12}^{+ 0.13}$ keV at the end of the extended emission phase. We also estimated the radii of the black body emitting region which varied from 0.11$\pm 0.01$ to 0.23$\pm0.03$ km during the extended tail in an irregular manner (see Figure~\ref{Aspecvar}).  The spectral best fits to the different time segments are shown in Figure~\ref{Aspecfits}. Adding a power law to the blackbody does not provide any improvement to the reduced chi-square of the fits. We tested the temporal evolution of the flux and black body temperature using the power law model. This model resembles the general decline trend even though it yields poor fit statistics (reduced chi-square values greater than three). The power law indices for the flux and temperature decay were obtained to be $-0.568\pm0.018$ and $-0.182\pm0.014$, respectively. Note that these indices matched closely with the reported values for other bursts from various sources \citep{Ibrahim2001, Lenters2003, Feroci2003} and also with the theoretical expectations \citep{Lenters2003, Thompson2002}.

It is important to emphasize the fact that the spectrum of the second segment after the burst (Tail 2) was fitted with a single black body model with statistically acceptable $\chi^2_{\nu} = 0.99$ for 60 dof. However there were clear signatures of spectral intensity deficits around 4.5 keV and 6.5 keV (see Figure~\ref{Aspecfits}). We modeled the spectrum of this segment with a combined model of a black body function and two (Gaussian) absorption lines. This model improves the fit statistics, $\chi^2_{\nu} = 0.66$ for 56 dof, and yields the centroid energies of 4.56$\pm$0.13 and 6.70$_{-0.17}^{+0.15}$ keV for the two absorption features.  

Next, we needed to check whether the addition of Gaussian absorption lines were statistically significant. In general, F-test is employed to test the statistical significance of the addition of parameters to an existing model. However, in the case of Gaussian line, this test is not applicable as clearly described in \citet{Protassov2002}. They pointed out that F-test is applicable only when the null values of the additional parameters are not coinciding with any boundary of the possible values of the parameters which is not the case for Gaussian models. Hence, we followed a prescription put forward by \citet{Protassov2002} to test the significance of adding a Gaussian component. First, we fit the data with a seed model (black body) and a test model (black body + two Gaussian line components) and calculated the observed F-statistic value. However, the distribution of F-statistic in our case is unknown. Therefore, we performed extensive simulations in order to determine the F-statistic distribution. We assumed the best-fit seed model to be the intrinsic model describing the spectrum, i.e., our null hypothesis was that the seed model is sufficient to describe the spectra and no additional parameters were required. Using the best-fit parameters of the seed model, we then generated 2000 simulated spectra using the {\tt fakeit} tool in XSPEC. Each simulated spectrum was then fitted with the seed and the test model and in each case the F-statistic value was calculated; from the resulting 2000 F-statistic values the F-distribution was constructed.
Our null hypothesis is that no line is present and alternative hypothesis states that some line components are required to describe the model sufficiently well. The null hypothesis probability is then given by the probability of obtaining F-statistic value equal to or larger than the measured F-statistic value purely due to noise. We obtained this probability by integrating the F-distribution above the measured value and call this null hypothesis probability p-value. If the null hypothesis probability is less than a certain threshold, in this case $2.7\times10^{-3}$ ($3 \sigma$) we reject the null hypothesis in favor of the alternate hypothesis. This means that if the null hypothesis probability is less than the threshold value we conclude that the test model is favored relative to the seed model.

From our simulations, the probability that the improvement of the chi-square on adding two Gaussian absorption lines to a single black body component being due to noise only was obtained to be less than $5\times10^{-4}$. This means that two absorption lines are required to describe the spectrum corresponding to the segment Tail 2 well. The spectrum corresponding to Tail 3 also shows signatures of absorption features, however from our simulations we obtained the null hypothesis probability for the measured F-statistic to originate purely from noise as $4.7\times10^{-3}$. For the spectra during the later part of the tail adding a single or double absorption lines does not significantly improve the model fitting. It should be noted that, the segment Tail 6 corresponded to a lower intensity burst and interestingly the black body temperature corresponding to this segment shows a significant increase and the black body radius shows a dip as also observed for the primary burst.

\subsubsection{Event B}

\begin{table*}[H!ta]
\scriptsize
\centering
\caption{Table displaying the best-fit spectral parameters for event B \label{SpecresultsB}}
\begin{tabular}{lccccccc}
\hline
Segment & Bestfit & kT & Radius & \multicolumn{2}{c}{Line} &  Flux$^a$ & $\chi_{\nu}^2$ (dof)\\
     & model & (keV) & (km) & E (keV) & Norm &$10^{-9}$ ergs/s/cm$^2$ & \\
\hline
\textbf{Burst} & \pbox{20cm}{PL$^b$ \\ 0.42$_{- 0.14}^{+ 0.13}$}
& ... & ... & ... & ... &  25.412$_{- 1.978}^{+ 1.971}$ &  0.75 (10)\\
\hline
\textbf{Gray 1} & BB
&  6.65$_{- 0.44}^{+ 0.50}$ &   0.28$_{- 0.02}^{+ 0.02}$ & ... & ... &  9.426$_{- 0.600}^{+ 0.598}$ &  1.57 (20)\\
\hline
\textbf{Gray 2} & BB+Abs+Abs
&  4.76$_{- 0.22}^{+ 0.24}$ &   0.35$_{- 0.02}^{+ 0.02}$ &     &     &  4.243$_{- 0.231}^{+ 0.234}$ &  1.99 (24)\\
Abs1 &  &  &  & 6.56$_{- 0.19}^{+ 0.19}$  & -0.02245$_{-0.00455}^{+0.00452}$ &  &  \\
Abs2 &  &  &  & 10.54$_{- 0.23}^{+ 0.23}$ & -0.01565$_{-0.00349}^{+0.00349}$ &  &  \\
\textbf{Gray 2} & BB+Abs+Abs+Ga
&  4.75$_{- 0.28}^{+ 0.31}$ &   0.32$_{- 0.02}^{+ 0.02}$ &     &     &  4.140$_{- 0.241}^{+ 0.245}$ &  1.35 (22)\\
Abs1 &  &  &  & 6.54$_{- 0.26}^{+ 0.26}$  & -0.01515$_{-0.00486}^{+0.00477}$ &  &  \\
Abs2 &  &  &  & 11.14$_{- 0.41}^{+ 0.45}$ & -0.01463$_{-0.01139}^{+0.00476}$ &  &  \\
Ems1 &  &  &  & 13.42$_{- 0.70}^{+ 0.48}$ & 0.02450$_{-0.00624}^{+0.01142}$ &  &  \\
\hline
\textbf{Gray 3} & BB+Ga
&  5.09$_{- 0.87}^{+ 0.50}$ &   0.16$_{- 0.02}^{+ 0.02}$ &     &     &  1.590$_{- 0.140}^{+ 0.143}$ &  0.76 (15)\\
Ems1 &  &  &  & 13.65$_{- 0.48}^{+ 0.51}$ & 0.01120$_{-0.00307}^{+0.00308}$ &  &  \\
\hline
\textbf{Osc} & BB+Ga
&  3.65$_{- 0.14}^{+ 0.14}$ &   0.19$_{- 0.01}^{+ 0.01}$ &     &     &   0.674$_{- 0.021}^{+ 0.022}$ &  1.45 (52)\\
Ems1 &  &  &  & 13.89$_{- 0.15}^{+ 0.15}$ & 0.00644$_{-0.00077}^{+0.00084}$ &  &  \\
\hline
\textbf{Tail 1} & BB+Abs+Abs
&  2.48$_{- 0.07}^{+ 0.07}$ &  0.24$_{- 0.01}^{+ 0.01}$  &     &     &   0.168$_{- 0.004}^{+ 0.005}$ &  1.69 (58)\\
Abs1 &  &  &  & 4.15$_{- 0.07}^{+ 0.07}$  & -0.00144$_{-0.00023}^{+0.00022}$ &  &  \\
Abs2 &  &  &  & 6.63$_{- 0.11}^{+ 0.11}$  & -0.00110$_{-0.00019}^{+0.00019}$ &  &  \\
\textbf{Tail 1} & BB+Abs+Abs+Ga
&  2.22$_{- 0.08}^{+ 0.08}$ &  0.30$_{- 0.02}^{+ 0.02}$  &     &     &   0.165$_{- 0.004}^{+ 0.004}$ &  1.36 (56)\\
Abs1 &  &  &  & 4.15$_{- 0.06}^{+ 0.06}$  & -0.00191$_{-0.00028}^{+0.00027}$ &  &  \\
Abs2 &  &  &  & 6.44$_{- 0.11}^{+ 0.12}$  & -0.00152$_{-0.00026}^{+0.00025}$ &  &  \\
Ems1 &  &  &  & 12.97$_{- 0.23}^{+ 0.23}$ & 0.00067$_{-0.00014}^{+0.00014}$ &  &  \\	
\hline
\textbf{Tail 2} & BB
&  2.18$_{- 0.07}^{+ 0.07}$ &  0.21$_{- 0.01}^{+ 0.01}$ & ... & ... &  0.083$_{- 0.003}^{+ 0.003}$ &  1.74 (63)\\
\hline
\textbf{Tail 3} & BB
&  1.78$_{- 0.07}^{+ 0.07}$ &  0.26$_{- 0.01}^{+ 0.01}$ & ... & ... &  0.055$_{- 0.003}^{+ 0.003}$ &  1.41 (63)\\
\hline
\textbf{Tail 4} & BB
&  1.83$_{- 0.08}^{+ 0.09}$ &  0.24$_{- 0.02}^{+ 0.02}$ & ... & ... &  0.051$_{- 0.003}^{+ 0.003}$ &  1.35 (63)\\
\hline
\textbf{Tail 5} & BB+PL
&  1.49$_{- 0.07}^{+ 0.08}$ &  0.34$_{- 0.03}^{+ 0.03}$ & ... & ... &  0.053$_{- 0.003}^{+ 0.016}$ &  1.12 (61)\\
\hline
\textbf{Tail 6} & BB+PL
&  1.51$_{- 0.08}^{+ 0.09}$ &  0.33$_{- 0.03}^{+ 0.03}$ & ... & ... &  0.051$_{- 0.007}^{+ 0.020}$ &  1.27 (61)\\
\hline
\textbf{Tail 7} & BB+PL
&  1.53$_{- 0.08}^{+ 0.09}$ &  0.30$_{- 0.02}^{+ 0.02}$ & ... & ... &  0.039$_{- 0.002}^{+ 0.002}$ &  1.10 (61)\\
\hline
\textbf{Tail 8} & BB+PL
&  1.48$_{- 0.09}^{+ 0.10}$ &  0.30$_{- 0.03}^{+ 0.03}$ & ... & ... &  0.033$_{- 0.002}^{+ 0.002}$ &  0.96 (61)\\
\hline
\end{tabular}
\begin{footnotesize}
\\
\begin{flushleft}
$^a$ Unabsorbed flux in 2.5-30 keV in units of $10^{-9}$ ergs/s/cm$^2$.\\
$^b$ The best-fit powerlaw index in quoted below.\\
All parameter errors quoted here $1 \sigma$ confidence errors.
\end{flushleft}
\end{footnotesize}
\end{table*}

\begin{figure}[H!tba]
\centering
\includegraphics[width=8cm]{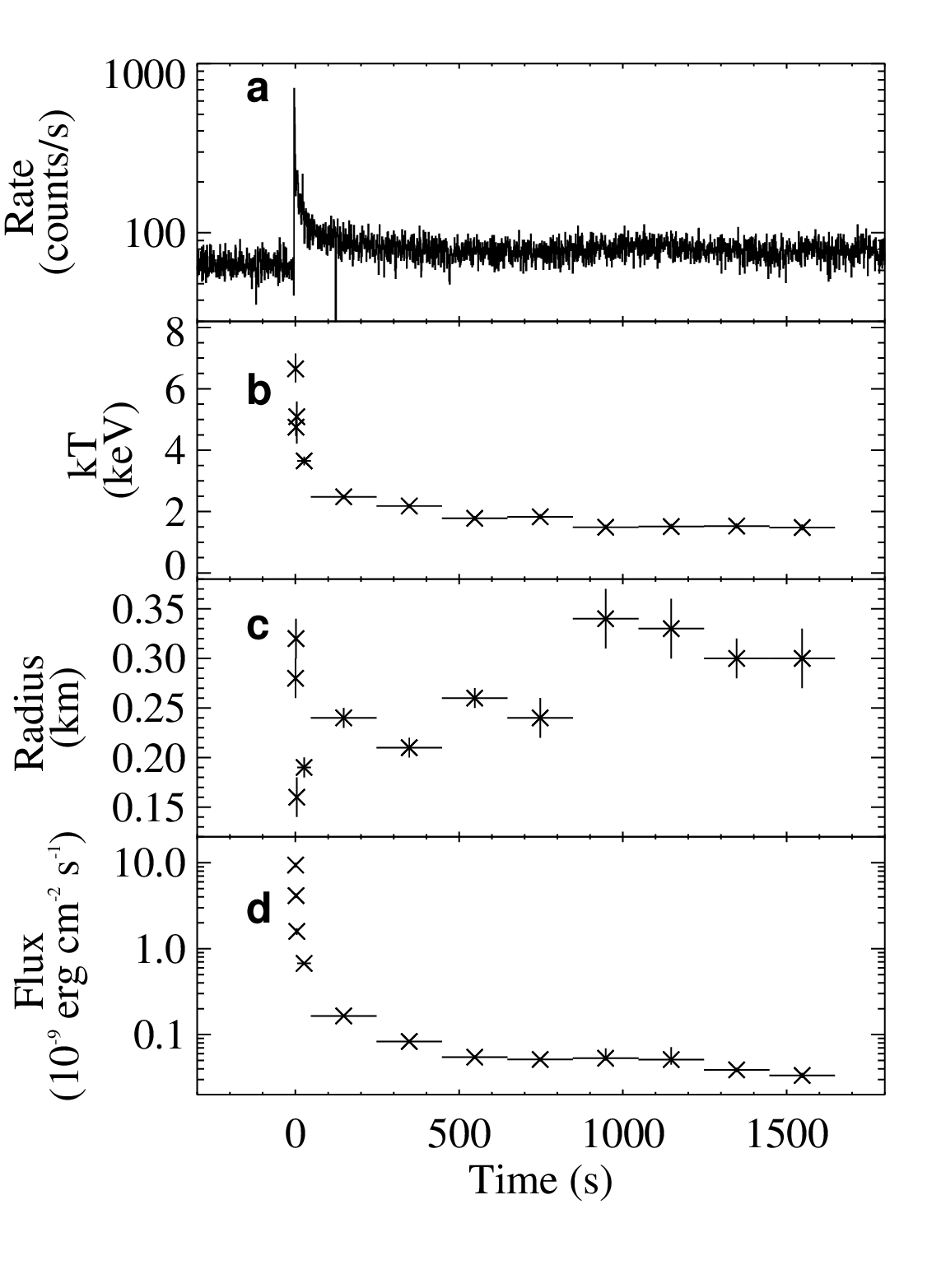}
\vspace*{-1.0cm}
\caption{Variation of best-fit spectral parameters in the energy range 2.5-30 kev during the extended tail corresponding to event B. (a) the source intensity of the event with 1s binning. (b) The evolution of the blackbody temperature. (c) The corresponding blackbody radius variation. (d) The variation of the 2.5-30 keV unabsorbed flux during the tail segments. All errors shown here are $1 \sigma$ errors on the parameters (the errors on the flux values are relatively too small to discern in the plot). \label{Bspecvar}}
\end{figure}

\begin{figure*}[H!tba]
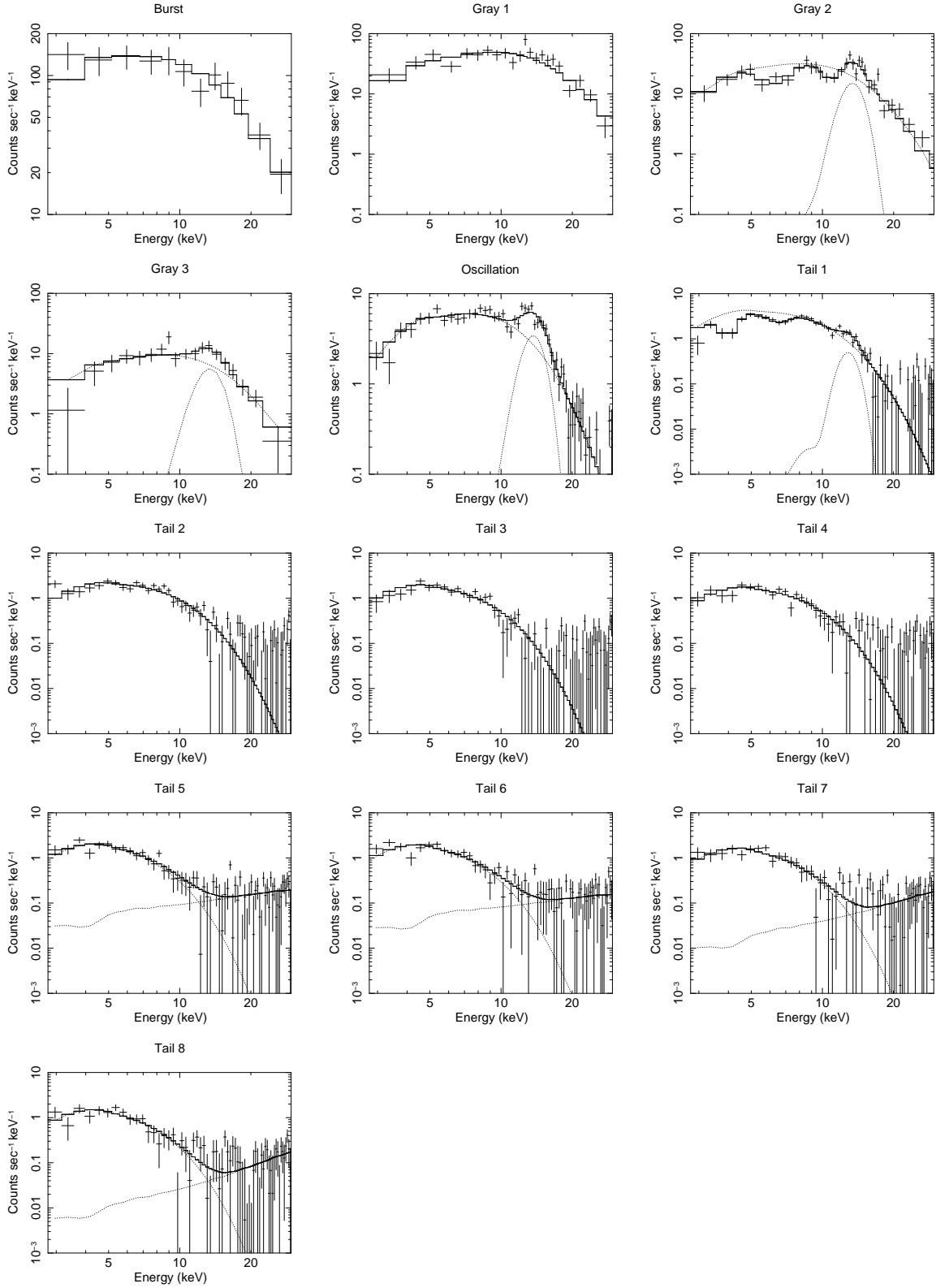

\centering
\begin{tabular}{ccc}
\includegraphics[width=4cm, angle=-90]{fig6_1.ps} &
\includegraphics[width=4cm, angle=-90]{fig6_2.ps} &
\includegraphics[width=4cm, angle=-90]{fig6_3.ps} \\
\includegraphics[width=4cm, angle=-90]{fig6_4.ps} &
\includegraphics[width=4cm, angle=-90]{fig6_5.ps} &
\includegraphics[width=4cm, angle=-90]{fig6_6.ps} \\
\includegraphics[width=4cm, angle=-90]{fig6_7.ps} &
\includegraphics[width=4cm, angle=-90]{fig6_8.ps} &
\includegraphics[width=4cm, angle=-90]{fig6_9.ps} \\
\includegraphics[width=4cm, angle=-90]{fig6_10.ps} &
\includegraphics[width=4cm, angle=-90]{fig6_11.ps} &
\includegraphics[width=4cm, angle=-90]{fig6_12.ps} \\
\includegraphics[width=4cm, angle=-90]{fig6_13.ps} 
\end{tabular}
\caption{Spectral fit of segments corresponding to event B. The label at the top of each panel denotes the segment that the corresponding spectrum was generated from. Each panel also show the best fit spectral model in the 2.5-30 keV band. \label{Bspecfits}}
\end{figure*}

We divided $\sim 1600$ s long tail into time segments of 200 s. As described for event A, we again extracted spectra from each time segment as well as for the leading burst, gray and oscillation intervals, and modeled them with thermal and non-thermal emission models. In Table~\ref{SpecresultsB} we list the details of the spectral fit results for event B and in Figure~\ref{Bspecvar} we present the evolution of best-fit spectral parameters during this event. The burst spectrum in this case is well described with a non-thermal single power law model (Figure~\ref{Bspecfits}). The fit corresponded to $\chi^2_{\nu}$ of 0.75 for 10 dof. A single black body model fit to the burst spectrum is unacceptable with $\chi^2_{\nu}$ of 2.17 for 10 dof. The 2.5$-$30 keV unabsorbed flux for the burst is $25.4 \pm 2.0 \times 10^{-9}$ erg s$^{-1}$ cm$^{-2}$ which corresponded to a burst luminosity of $3.7\times10^{37}$ erg s$^{-1}$ assuming isotropic emission.

The spectra during the transitional (gray) intervals and the extended tail were best fitted using a black body model, but in some cases along with single or multiple absorption and emission lines. The black body temperature exhibited a smooth cooling trend with its value varying from 6.65$_{- 0.44}^{+ 0.50}$ to 1.48$_{- 0.09}^{+ 0.10}$ keV. The corresponding radii of the black body component varied from $0.16\pm0.02$ to $0.34\pm0.03$ km and showed slightly different behavior; it remained around 0.23 km during the first half of the tail, and around 0.32 km during the rest (see Figure~\ref{Bspecvar}). For the gray region, the 2.5-30 keV unabsorbed flux ranged from $9.25\pm0.60\times 10^{-9}$ to $1.59\pm0.14\times 10^{-9}$ erg s$^{-1}$ cm$^{-2}$ and during the extended tail, the flux decayed from 0.165$\times 10^{-9}$ to 0.033$\times 10^{-9}$ erg s$^{-1}$ cm$^{-2}$. Again assuming isotropic emission, we found the corresponding luminosity varying from $2.46\times10^{35}$ and $0.48\times10^{35}$ erg s$^{-1}$ during extended tail emission phase. Here also, we fitted the temporal evolution of flux and black body temperature with the power law model, and obtained consistent power law indices of $-0.720 \pm 0.006$ and $-0.190 \pm 0.002$, respectively, again despite the rather poor fit statistics. Next, we give a detailed description of the spectral modeling of the segments falling in the gray, oscillation and extended tail intervals, in particular when more complex spectral models were needed to describe their spectra. The spectra corresponding to the various time segments along with the best fit models are shown in Figure~\ref{Bspecfits}. 

The spectrum of the Gray 1 segment, just following the burst, was described best using  a single black body model with temperature of 6.65 keV.
Note that this spectral shape is quite similar to the burst spectrum in event A.
The following segment, Gray 2, was much more complex and we modeled using a black body with a combination of absorption and emission line features: Two absorption lines at 6.56 and 10.54 keV on top of the base black body model fitted the spectra yields $\chi^2_{\nu}$ of 1.99 for 24 dof. Adding an emission line at 13.42 keV to this model improves $\chi^2_{\nu}$ to 1.35 for 22 dof (see Figure~\ref{Bspecfits}). We then examined the statistical significance of addition of these lines via the same technique as described in the previous sub-section for event A. In this case, our seed model was a single black body as before and the test model was black body + two absorption lines + an emission line. We performed the simulation taking the best-fit single black body model as the intrinsic model and the fitted the simulated spectra using the seed and test models. The F-statistic values for the seed and test model were calculated for each simulated spectra and the F-distribution was constructed. We compared the measured F-statistic value against the constructed F-distribution and obtained the p-value to be less than $5\times10^{-4}$, hence strongly favoring the addition of the Gaussian line components. 
We also checked the significance of adding the emission line along with the black body and the two absorption lines. Again from our simulations, we find that the null hypothesis probability taking the black body + two absorption lines as the seed model and the black body + two absorption lines + an emission line as the test model turns out to be $1.6\times10^{-3}$.
The spectrum corresponding to the next segment Gray 3 did not show any distinct absorption features but was well described with a black body along with an emission line at 13.65 keV (Figure~\ref{Bspecfits}). As before, we calculated the significance of adding this line from simulation by taking the single black body as the seed model and the black body with an emission line as the test model. We obtained the p-value (see previous section) as $7\times10^{-4}$ rendering the addition of the Gaussian parameters as highly significant.	
The succeeding segment Oscillation (Osc) displayed a similar spectrum and was best fitted using black body model with an emission line at 13.89 keV (Figure~\ref{Bspecfits}). We obtained the null hypothesis probability i.e, the p-value to be less than $5\times10^{-4}$, thereby strongly favoring the test model.

The extended tail began following the Oscillation segment. The first segment, Tail 1, showed a complex spectrum with multiple line features. A black body model with absorption lines at 4.15 and 6.63 keV described the spectrum well giving a $\chi^2_{\nu}$ of 1.69 for 58 dof. Adding an emission line at 12.97 keV, improves the fit significantly; $\chi^2_{\nu}$ 1.36 for 56 dof (see Figure~\ref{Bspecfits}). Again from the simulation, considering the single black body as the seed model and the black body + two absorption lines + an emission line as the test model, we found the p-value to be less than $5\times10^{-4}$ favoring the addition of the line components. We also performed a comparative significance testing of the two best-fit models for Tail 1 segment described here, that is to say we also tested the significance of adding the emission line on top of the black body with the two absorption lines. In this case, the p-value that the seed model (black body+ 2 absorption lines) better describes the data as compared to the test model (black body+ 2 absorption lines + an emission line) was computed to be $7\times10^{-4}$. 

Following this, the tail spectra became relatively simpler and was primarily described using a single black body model. The spectra of some of the segments, such as, Tail 2 and even some of the later spectra (e.g., Tail 6) showed indications of certain absorption features, particularly around 4 keV, but none of Gaussian line features were sufficiently significant as obtained from our simulations. During the later part of the tail, from the spectrum of the segment Tail 5, a substantial excess was observed in the high energies above 10 keV. We modeled the high energy excess using a power law and adding such a component effectively improved the fits (see last four panels of Figure~\ref{Bspecfits}). Upon the addition of a power law component, $\chi^2_{\nu}$ improved from 2.07 (63 dof) to 1.15 (61) for Tail 5, 1.91 (63) to 1.27 (61) for Tail 6, 1.64 (63) to 1.10 (61) for Tail 7 and 1.40 (63) to 0.96 (61) for Tail 8 though the spectral parameters of the power law component could not be well-constrained. The addition of this power law component to the spectral fits of the preceding segments did not perceptibly improve the fit statistics.
Here also we investigated the statistical significance of the addition of extra power law component on top of the black body component. For these spectra, we constructed and fitted the simulated spectra using the black body as the seed model and the black body + power law as the test model, and computed the F-distribution. We observed that, in all cases for the last four tail segments, the null-hypothesis probability or p-value was less than $5\times10^{-4}$ strongly favoring the addition of the power law component. Such high energy excess was previously reported for this source by \citet{denHartog2008, Trumper2010, Trumper2013}. 

Finally, it is interesting to note here that for both in event A and B, the transient line features appear at similar times, immediately following the burst, and then disappear as the tail decays. For both events, we observed absorption line features at $\sim 4$ keV and $\sim 6.5$ keV just after the intense burst emission. Additionally, a 13$-$14 keV emission line was present in the spectra of event B which was previously reported by \citet{Gavriil2011}.

\subsection{Temporal Analysis}\label{Temp}
\subsubsection{Variation in Pulsed Properties}
As we presented in Section \ref{Tail}, periodic intensity modulations at the spin frequency of the underlying source are clearly visible in the tail emission of both events. To better understand burst-induced changes on the persistent emission of 4U 0142+61, we also investigated the time evolution of the pulsation amplitudes during the extended tails of these bursts, and compared them to those during the pre-burst values. For this purpose, we considered the same segments during the extended tail used for the spectral analysis; again the possible weak burst features during these tail segments were not excluded as they could not be significantly distinguished from noise. For each segment, the data was folded according to the spin ephemeris given in \citet{Gavriil2011} where $\nu$ and $\dot{\nu}$ were 0.1150920955(12) Hz and $-2.661(9)\times10^{-14}$ Hz s$^{-1}$, respectively. Apart from this long term timing solution, they inferred a glitch at an epoch of $53809.185840$ MJD. The glitch affects the long term phase coherent timing solution as the frequency and hence the phase takes sufficient time to recover back the original pre-glitch timing behavior. Therefore, we use the timing solution for frequency evolution also provided by \citet{Gavriil2011} which comprises of a frequency model during the glitch along with the pre-glitch evolution, that is: 
\begin{equation}
\nu = \nu_0(t) + \Delta \nu + \Delta \nu_d e^{-(t-t_g)/\tau_d} + \Delta \dot{\nu} (t-t_g)
\end{equation}
and $\nu_0(t)$ is the pre-glitch frequency evolution given as:
\begin{equation}
\nu_0(t) = \nu(t_0) + \dot{\nu} (t-t_0) 
\end{equation}
where $\Delta \nu$ is the instantaneous frequency jump at the onset of the glitch, $\Delta \nu_d$ is the change in frequency in a timescale of $\tau_d$ during the recovery, $t_g$ is the glitch epoch and $\dot{\nu}$ is the post-glitch modification to the frequency derivative of the long term timing solution. The values of the parameters used for the final frequency evolution model were $-1.27(17)\times10^{-8}$ Hz for $\Delta \nu$, $-3.1(1.2)\times10^{-16}$ Hz s$^{-1}$ for $\Delta \dot{\nu}$, $2.0(4)\times10^{-7}$ Hz for $\Delta \nu_d$ and $17.0(17)$ days for $\tau_d$ \citep{Gavriil2011}. We then constructed the pulse profiles in the energy range 2--10 keV over each of spectral investigation time segments during the extended tails. Next, these profiles were modeled with a sinusoidal model along with four harmonics following \citet{Dib2007}. From the modeling, we computed the rms fractional amplitudes of the pulsation and its variation prior to the bursts and during the decaying tails. In both cases, the pulsations are significantly enhanced following the burst episode, as also reported by \citet{Gavriil2011}. Since the extended light curves of tails show significant variation with energy and their spectra changes with time, we also examined the energy dependence of the pulsed amplitude evolution by repeating the same pulse profile modeling technique in two constituting energy ranges 2--5 and 5--10 keV. 

\begin{figure}[H!tba]
\centering
\includegraphics[width=8cm]{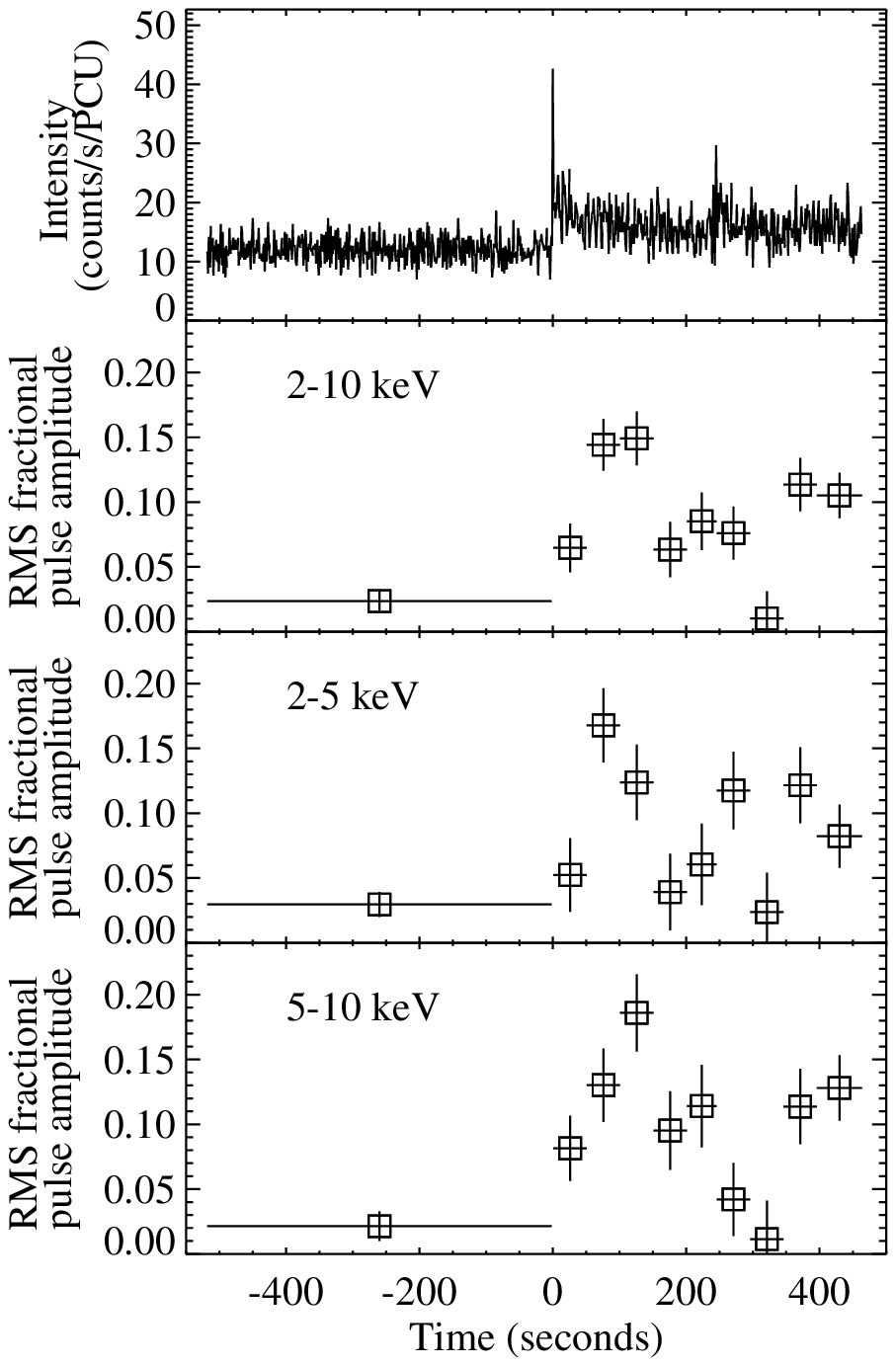}
\caption{The variation of the rms fractional amplitude of pulsation with intensity during the burst and the burst and extended tail emission during event A i.e., the observation on June 25, 2006. The second, third and fourth panels show the variation of the rms fractional amplitude in the 2-10, 2-5 and 5-10 keV energy ranges respectively. All parameter errors quoted here 1$\sigma$ confidence errors.\label{Apulsevar}}
\end{figure}

For event A, the average pre-burst rms fractional pulsation amplitude is about 2\%, which is in line with the fact that 4U 0142+61 is the magnetar with the lowest rms pulsed fraction \citep{Patel2003}. The rms pulsed amplitude following the leading burst went up to 15\% in 50 s and remained at this level for about 100 s (see the second panel of Figure~\ref{Apulsevar}). It then declines down to about 7\% and remains at this level for 150 s interval which includes the time of the weaker burst in the extended tail. Interestingly, the rms pulsed amplitude briefly drops back to the pre-burst level, and then remains at about 10\% level until the end of extended burst tail (Figure~\ref{Apulsevar}). The evolution of the rms fractional amplitude in the 2--5 keV and 5--10 keV energy ranges are a bit more complicated, but to some extent resemble the behavior in 2--10 keV band (the third and fourth panels of Figure~\ref{Apulsevar}, respectively.)

\begin{figure}[H!tba]
\centering
\includegraphics[width=8cm]{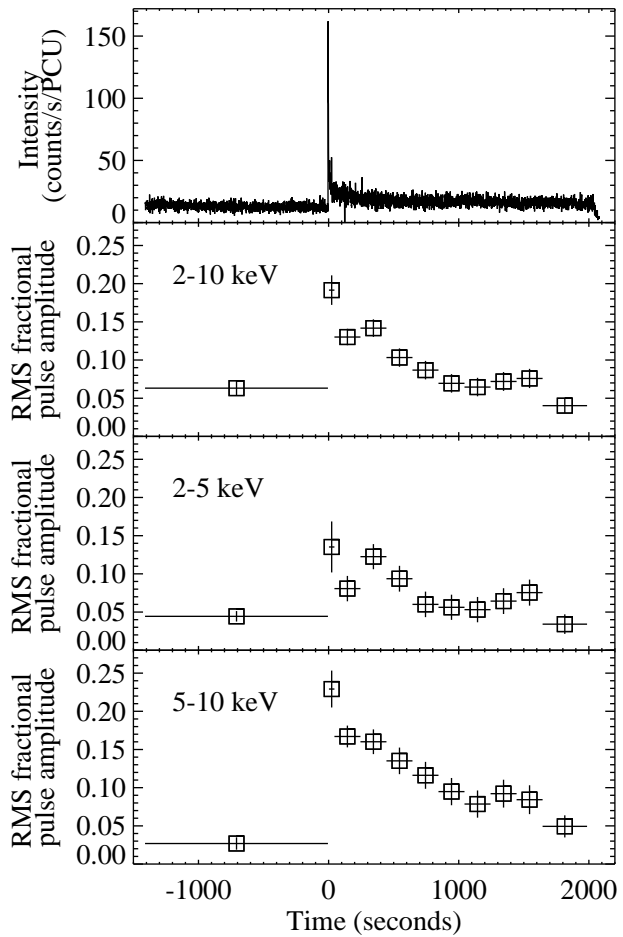}
\caption{The variation of the rms fractional amplitude of pulsation with intensity during the burst and the burst and extended tail emission during event B i.e., the observation on February 7, 2007. The second, third and fourth panels show the variation of the rms fractional amplitude in the 2-10, 2-5 and 5-10 keV energy ranges respectively. All parameter errors quoted here 1$\sigma$ confidence errors.\label{Bpulsevar}}
\end{figure}

For event B, the pre-burst pulsation amplitude was higher, $\sim$6\%. It shows an abrupt enhancement in conjunction with the leading burst to about 20\% (the second panel of Figure~\ref{Bpulsevar}). The pulsed amplitude then declines smoothly to the pre-burst level in about 800 s, and remains nearly constant at this level (or with a marginal increase) for the next 800 s (Figure~\ref{Bpulsevar}). The energy resolved rms pulsed amplitude of this event is intriguing: In the 5--10 keV band, it goes up to about 23\% from 3\%, and then decays steadily down to about 10\% until the end of the whole tail (the fourth panel of Figure~\ref{Bpulsevar}). In the 2--5 keV band, however, the initial enhancement is not as large, rising from 4\% to 13\%, and decayed back to the pre-burst level in about 600 s.  The rms amplitude in the lowest investigated energy regime remains constant for about 600 s and then shows an increasing trend for about 400 s, before it rapidly drops back to the pre-burst level. It is important to note that characteristic rms pulsed amplitude variations seen in the 2--10 keV band during the first 600 s and the last 800 s of the extended tail are mostly due to variations in the 2--5 keV band (Figure~\ref{Bpulsevar}).

\subsubsection{Rotational Phases of the Events}

\begin{figure}[H!ba]
\centering
\begin{tabular}{c}
\includegraphics[width=9cm]{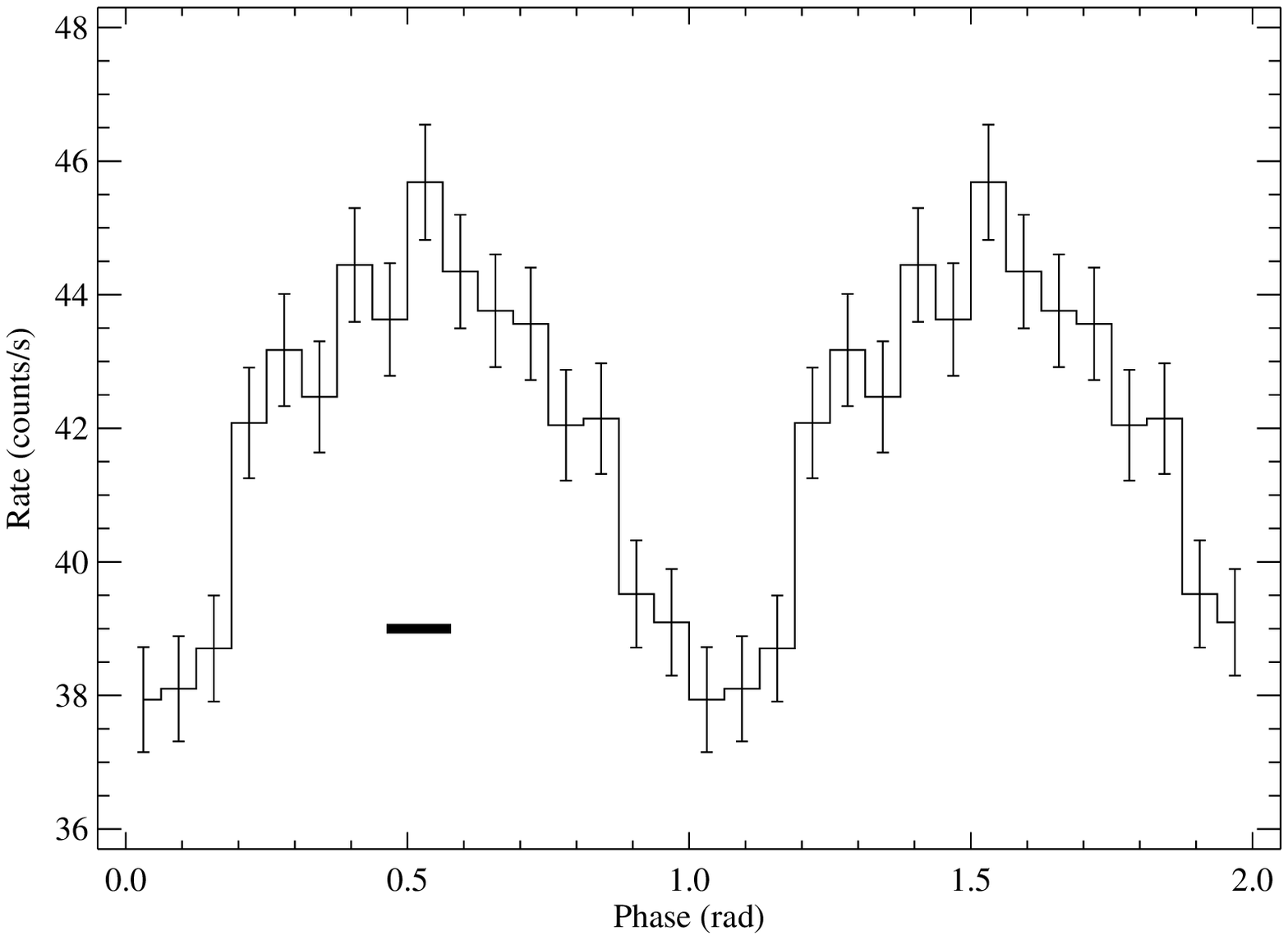} \\
\includegraphics[width=9cm]{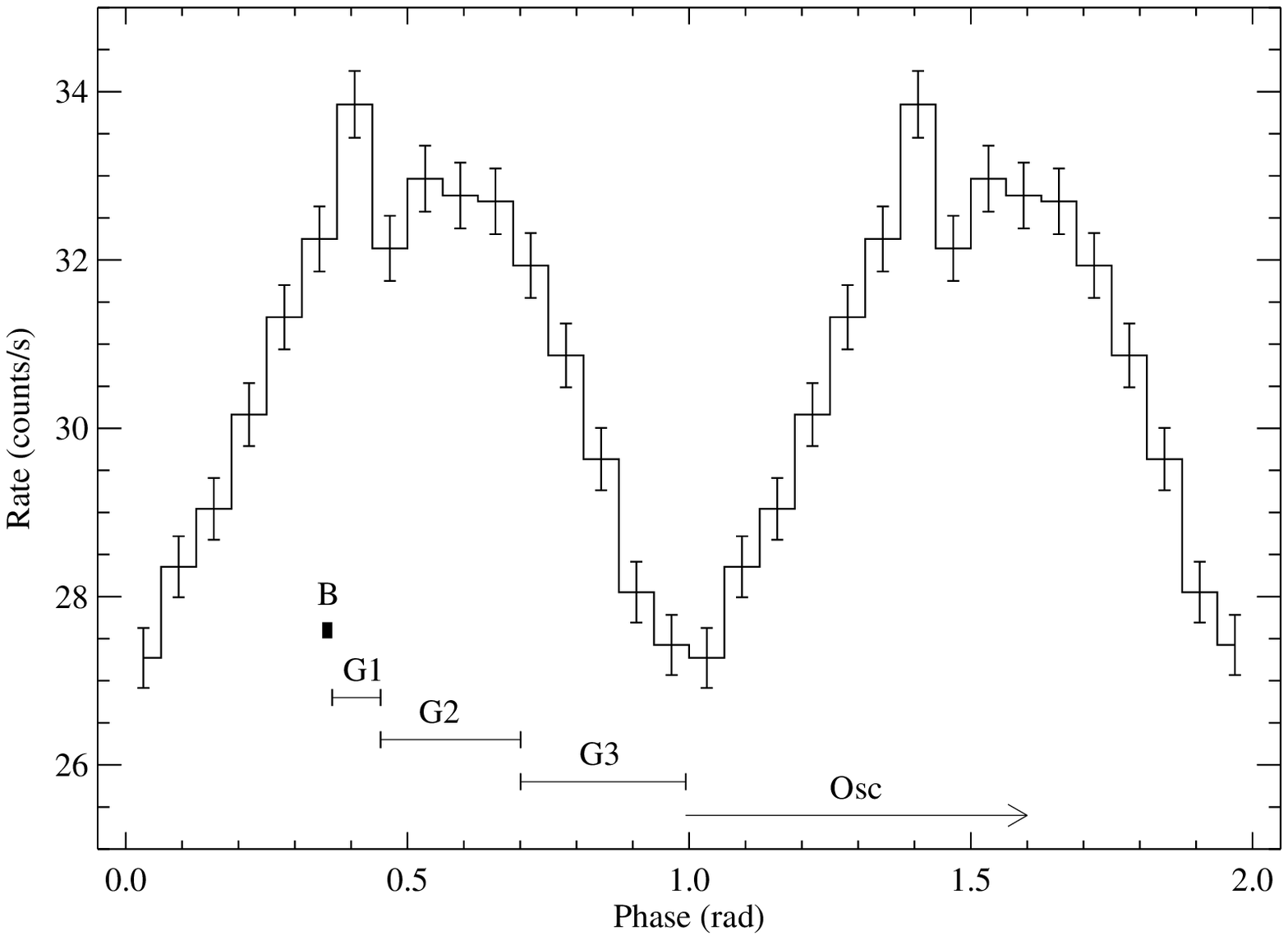}
\end{tabular}
\caption{Top panel: pulse profile constructed from the pre-burst data corresponding to event A. The black bar shows the phase interval of the burst for the event A. Bottom panel: pulse profile constructed from the pre-burst data corresponding to event B. The black bar marked B shows the phase interval of the burst for the event B, the intervals marked G1, G2 and G3 denotes the phase intervals of the segments Gray 1, Gray 2 and Gray 3 respectively. The arrow marked Osc denotes the start of the Oscillation region. All errors shown here 1$\sigma$ confidence errors.\label{phasealign}}
\end{figure}

We determined the rotational phases of both events to uncover any link between the burst emission and persistent X-ray output of 4U 0142+61. We used the spin ephemerids of \citet{Gavriil2011} to calculate the phases of the bursts. For event B, we also calculated the phases of the short ``gray" episodes, which took place in transition from the burst to the episode of clearly visible ``oscillation" regime. We find that the burst corresponding to event A spans between the spin phases 0.46 and 0.58 (see top panel of Figure~\ref{phasealign}) which is entirely aligned at the maximum intensity interval of the pulse profile. For event B, the short intense burst spans a much shorter spin phase interval from 0.35 to 0.37, that also coincides with a peak intensity interval of the underlying pulse profile which is present even after removing the burst time interval. The burst interval is followed by the Gray 1 interval which ends at the spin phase of 0.45, then by the Gray 2 segment which ends at 0.70. The succeeding segment, Gray 3 continues until the rotation phase of 0.99 after which the oscillation regime starts that corresponds to a duration of about 42 s comprising multiple spin cycles. The spin phase of these various segments in event B are presented in the bottom panel of Figure~\ref{phasealign}. Note the fact that the alignment of bursts with the peak of pulse profile were also reported by \citet{Gavriil2011}.

\subsubsection{Search for QPOs}

Quasi-periodic oscillations (QPOs) have been detected during the giant magnetar flares \citep{Israel2005, Strohmayer2005} and in some short recurrent bursts \citep{Huppenkothen2014a, Huppenkothen2014b}. We, therefore, searched the power spectra of the extended tails of the two events for the presence of QPOs. First, we segmented the extended tails into 1 second intervals and constructed the Leahy-normalized power spectra \citep{Leahy1983} in 2.0-14 keV energy band from them using a Nyquist frequency of 1024 Hz. If a power is more than $3\sigma$ significant i.e., the probability that this power to originate purely due to noise is less than 0.27\%, then we call it a signal detection. If the number of trials (the number of frequency bins searched) is taken into consideration, we find no significant signal in the acquired power spectra. We performed the same QPO search analysis over the energy bands 2.0-4.5, 4.5-8.5 and 8.5-14 keV but still did not find any detection that satisfies our criterion. 

In order to reduce the noise in individual power spectra, we generated the averaged power spectrum of these 1 s intervals and searched for QPOs in the frequency range from 10 Hz to 1020 Hz. We computed the chance probability of the maximum power in the relevant frequency range and found it to be greater than 0.0027 in all cases. We re-computed the chance probability and sigma level for the maximum power detected in the averaged power spectra for the two energy ranges 2.0-10.0 keV and 10.0-30.0 keV. Again it was a null detection when considered the number of trials. We obtained the significance of the maximum power during the tail of event A as 1.12$\sigma$ and 1.57$\sigma$ in the energy ranges 2--10 and 10--30 keV, respectively. For event B, the significances of the maximum detected power were obtained to be 0.06$\sigma$ and 0.10$\sigma$ for these two energy bands, respectively.

\section{Discussion}\label{Discussion}
We characterized the extended tails following the bursts from 4U 0142+61 reported in \citet{Gavriil2011} via thorough spectral and temporal analyses, and time evolution of its properties. The two extended burst tails we identified last longer than 463 and 1600 s, that are similar to the durations of the extended emission episodes observed from other magnetars \citep{Ibrahim2001, Lenters2003, Gogus2011, SasmazMus2015}. Below we discuss the implications of our detailed time-resolved spectral and temporal analysis results, and compare them with the broad spectral properties of other extended tails reported in literature. 

We showed that spectra of the extended tail episodes of both events were varied in time. Earlier, \citet{Gavriil2011} investigated X-ray spectra of these two events but they accumulated a spectrum from the entire duration of each event, therefore, overlooked the time variation. The burst corresponding to event A was described by a blackbody model with a temperature 5.6 keV which is consistent with the temperature reported by \citet{Gavriil2011} for the entire event duration. The subsequent extended tail spectra were also described with a thermal model and showed a smooth cooling trend; temperature decayed from 3.9 to 1.8 keV in about 400 s. Such time dependent spectral behavior was observed from the extended tail emission following bursts from a number of magnetars \citep{Ibrahim2001, Lenters2003, Gogus2011, SasmazMus2015}. For event B, the most adequate model to describe the burst spectrum was non-thermal; a power law with an index of 0.42. Such non-thermal behavior is also typical for magnetar bursts \citep{vanderHorst2012, Lin2013, SasmazMus2015}. Following the burst, the segments Gray and Oscillation were observed during which clear periodic modulations at 8.7 s could be seen. We found that the spectra corresponding to these regions were relatively more complex and displays distinct signatures of spectral line features. However, the continuum of the extended tail of event B has also exhibited a cooling trend from 2.5 to 1.5 keV in a slightly longer time scale, 1600 s. For other sources exhibiting extended tails as reported in the literature, the black body temperature declined in similar fashion. For SGR 1900+14, the temperature decreased from 2.5 to 1.6 keV for the 2001 April 28 burst extended tail and between 4.4 to 1.5 keV for the 1998 August 29 burst extended tail over time scales of few thousand seconds \citep{Lenters2003}. The temperature decayed from 3.01 to 1.73 keV, 2.37 to 1.69 keV and 3.23 to 2.20 keV over a few hundred seconds during the extended tails of events B, C, and D respectivey from the magnetar SGR J1550--5418 \citep{SasmazMus2015}. In the case of SGR 1806--20, the temperature decreased from 3.8 to 2.6 keV over ~500s during the extended tail corresponding to the event on 2004 June 22 and from 4.0 to 2.6 keV over about a thousand second during the extended tail corresponding to the event on 2004 October 17 \citep{Gogus2011}. In 4U 0142+61, the ``gray" regions precede the extended tail where the black body temperature was higher compared to the typical temperature observed during the onset of extended tails, and importantly, most of the prominent line features were observed during these segments.

Magnetar bursts are proposed to originate from small or large scale fracturing of the solid neutron star crust \citep{Thompson1995} or from magnetic reconnection \citep{Lyutikov2015}, in either case giving rise to a ``fireball" trapped in the magnetosphere, comprising of highly energetic charged pairs. Returning particles back onto the neutron star surface from this fireball that could not efficiently radiate away can heat up the stellar surface, therefore, can cause the extended tail emission. The cooling of the hence formed hot-spot results in the observed temperature decay trend. Moreover, during the burst, the high temperature reached could initiate thermonuclear burning releasing additional energy \citep{Ibrahim2001}. The fireball created during a burst event could compress the deep layers to trigger such a reaction. Such events can only take place if burst is sufficiently energetic resulting in a high fireball temperature and also if significant hydrogen is present as the burning fuel \citep{Ibrahim2001}. 
The time scale of the hydrogen burning process is consistent with the duration of the burst segment. The energy released in the burning of the hydrogen additionally heats up the crust  which can supersede the thermal energy injected into the crust directly from the fireball \citep{Ibrahim2001}. This means a large amount energy is released during the tail from multiple mechanisms. This phenomena is very much in accord with our observations where we find the energy released during the tail is quite high, being 1-2 order of magnitude larger the that released during the burst. Also the prolonged emission is consistent with the time taken for the heat to get released from the deep layers where the burning takes place. Thus thermonuclear burning may be another possible mechanism behind the long extended energetic tail emission.

We detected strong evidences of episodic spectral lines mainly at the onset of the extended tails. We have diagnosed the statistical significance of such lines through rigorous simulations in order to employ an unbiased approach to test the importance of adding the extra line components \citep{Protassov2002}. We detected an emission line at $\sim13$ keV and absorption lines at $\sim 4$, $\sim6.5$ and $\sim11$ keV with high significance. If the absorption features are caused by proton cyclotron resonance (see below), these lines correspond to 1.1$\times$10$^{15}$, 1.9$\times$10$^{15}$, and 3.1$\times$10$^{15}$ G, respectively (assuming a neutron star with mass of 1.4 M$_{\odot}$ and radius of 12 km). These magnetic field strengths are significantly higher than the inferred dipolar field strength of 1.3$\times$10$^{14}$ G, strengthening the idea that multi-polar surface magnetic field structures can be much more intense than the dipole field. 

Emission and absorption features at similar energies were previously reported in the burst spectra of a number of magnetar sources like 1E 1048.1--5937 \citep{Gavriil2002, Gavriil2006, An2014}, XTE J1810--197 \citep{Woods2005}, 4U 0142+61 \citep{Gavriil2011}, SGR 1806--20 \citep{Ibrahim2002, Ibrahim2003} and SGR 1900+14 \citep{Strohmayer2000}.
\citet{Strohmayer2000} claimed the detection of a 6.4 keV emission line during a SGR 1900+14 burst during a 0.3 s long segment when its spectrum was the hardest. They also reported the indication of a weak emission line at $\sim 13$ keV which they suggested to be a possible harmonic of the 6.4 keV feature. \citet{Gavriil2002} found the presence of an emission line feature at about 14 keV during the initial stages of the burst from 1E 1048.1--5937. A 5 keV absorption line was detected during a burst precursor from SGR 1806-20 by \citet{Ibrahim2002} and was attributed to proton cyclotron resonance in an ultra-strong magnetic field of a magnetar. An alternative possibility behind the origin of the line might be a red-shifted absorption line arising from an atomic transition like iron \citep{Strohmayer2000} but this is unlikely given the obtained line parameters. The observed flux and temperature corresponding to the line can not be explained by any existing theoretical model when the obtained red-shift is taken into account \citep{Ibrahim2002}. A 12.6 keV emission line, with a chance probability of $<4\times10^{-6}$, was discovered by \citet{Woods2005} during tail of a burst from XTE J1810--197. \citet{An2014} observed an emission line at $\sim13$ keV during the tail of a 1E 1048.1--5937 burst using {\it NuStar} data and this was the first detection by a non-{\it RXTE} instrument, ruling out instrumental effects. Note that these emission and absorption features were transient events during bursts and usually occurred during the segments when the spectrum was harder. 

One possible interpretation of the absorption lines at $\sim4.5$ or at $\sim6.5$ keV during the  4U 0142+61 extended tails is that they were generated due to the electron cyclotron resonance process but this is unlikely as this requires the magnetic field to be much lower ($\sim 10^{12}$ G) than the inferred dipolar field strength ($\sim10^{14}$ G). The other possibility behind these lines might be the ion or proton cyclotron resonance phenomenon. A trapped fireball is formed during a burst in both the crustal fracturing \citep{Thompson1995} and the magnetic reconnection \citep{Lyutikov2015} scenarios for bursts. Photons emitted from such a fireball interacts with the magnetosphere in presence of a strong multipolar magnetic field ($\sim 10^{15}$ G) giving rise to the proton cyclotron features. In this case the inferred magnetic field is quite close to the surface dipole magnetic field. As the temperature and flux decays, the proton cyclotron resonance process becomes inefficient and consequently the lines become weak or entirely absent in the late stages of the extended tail. This picture very much concurs with our observations of the absorption lines being transient and occurring at high intensities just following the burst. It is important to point out that the observed properties of the line features should be strongly dependent on the latitude of the emission site and also on the stellar rotation, both of which affect our viewing angle \citep{Thompson2002}. This might explain the presence and absence of the line features in two adjacent segments particularly in the gray regions of the event B.

\citet{Wang2006} observed this source using {\it Spitzer} in the mid-infrared wavelength band. The IR emission was best described by a multi-temperature thermal model indicating an extended disk origin. It was attributed to the reprocessing of incident X-rays coming from the central X-ray pulsar by the surrounding passive disk. \citet{Ertan2007} interpreted this observation as evidence of an active fallback disk with intrinsic disk emission due to viscous dissipation along with the reprocessing of the incident radiation. Existence of such a disk was predicted previously by \citet{Chatterjee2000} and \citet{Alpar2001}. As such a disk is present in 4U 0142+61, the $\sim6.5$ keV absorption line may have been caused due to absorption by the disk corresponding to the Fe K$\alpha$ transition \citep{Strohmayer2000}. As the source intensity changes during the burst, the incident flux and the reprocessing from the flared up disk varies which may in turn affect the presence of the absorption/emission line and thus can explain their transient nature. 

There has been only a few reported detections of absorption and emission lines in the burst or tail phases of emission of magnetars as mentioned before. Our observations that the spectral lines appear at the similar episodes during the tail for both events, where we have high intensity spectra, is intriguing. Such features, if indeed originating from cyclotron resonance, provides us with a unique and independent way of determining the stellar magnetic field and its properties. Moreover, a measurement of the gravitational redshift of these lines, most likely originating from near the stellar surface, can put significant constraints on the neutron star mass and radius. An instrument with better spectral capability than {\it RXTE} would be able to precisely characterize these line features.

It has been observed that the occurrence of the extended tail and its energetics are not closely related the energetics of the leading burst \citep{Gogus2011, SasmazMus2015}. We observed that the 2.5-30 keV isotropic energy emitted during the burst for event A was $2.3\times10^{36}$ erg which was followed by an extended tail with an isotropic energy in the same range of $4.9\times10^{37}$ erg. For event B, the total isotropic burst energy was $4.7\times10^{36}$ erg and that during the corresponding gray, oscillation and extended tail regions were found to be $29.4\times10^{36}$, $41.5\times10^{36}$ and $2.3\times10^{38}$ erg respectively. The ratio of the extended tail energy to the burst energy varies substantially between these burst events for this source. Possibly, a considerable fraction of the total energy released as a result of crustal fracturing or during a magnetic reconnection event could not be efficiently radiated away during the burst and was retained within the system, as also invoked by \citet{SasmazMus2015}. This residual energy reservoir could then power the emission during the extended tail over a prolonged period of time.

The temporal properties of 4U 0142+61 also exhibited interesting evolution during the extended emission phase. We observed that the burst induced enhanced pulsed amplitude in both events. It is interesting to note that the variation of the pulses amplitude, particularly for event A, is not correlated with the variation in the X-ray intensity and the pulsation remains high until into the extended tail, long after the burst occurrence. We also found that the spin phases of leading bursts in both events corresponded to the peak of the pulse profile. This might imply that the site of the leading burst is near the magnetic pole of the neutron star. As described in the implications of our spectral results, major fraction of the energy was deposited back to the system which heats up the star, and in turn this could result in the significant increase in the pulsation amplitude just following the burst and also during the later parts of the tail. Similar phenomena was observed for event D in \citet{SasmazMus2015} which had similar energetics (low energy bursts and more energetic long extended tails) much like the two events described in this work.	   

\section*{Acknowledgments}
We thank the referee for constructive comments and suggestions which have improved the paper. MC is supported by the Scientific and Technological Research Council of Turkey (T\"{U}B\.{I}TAK) through Research Fellowship Programme for International Researchers (2216).

{} 
\end{document}